\documentclass{sig-alternate-2013}

\permission{}
\usepackage{etoolbox}
\makeatletter
\patchcmd{\maketitle}{\@copyrightspace}{}{}{}
\makeatother

\usepackage{array}
\usepackage{multirow}
\usepackage{leqno}
\usepackage{graphicx}
\usepackage{url}
\usepackage{subfigure}
\usepackage{amssymb}
\usepackage{float}
\usepackage{marvosym}
\usepackage{epstopdf}
\usepackage{epsfig}
\usepackage{array}
\usepackage{amsmath}
\mathchardef\mhyphen="2D

\begin{document}


\title{Stories From the Past Web}
\numberofauthors{2}%
\author{
\alignauthor Yasmin AlNoamany,\\
 \affaddr{University of California, Berkeley }\\
 \affaddr{Berkeley, CA, USA}
 \email{yasminal@berkeley.edu}
\alignauthor
Michele C. Weigle, Michael L. Nelson\\
 \affaddr{Old Dominion University}\\
 \affaddr{Norfolk, VA, USA}
 \email{\{mweigle, mln\}@cs.odu.edu}
}

\maketitle

\begin{abstract}
Archiving Web pages into themed collections is a method for ensuring these resources are available for posterity. Services such as Archive-It exists to allow institutions to develop, curate, and preserve collections of Web resources. Understanding the contents and boundaries of these archived collections is a challenge for most people, resulting in the paradox of the larger the collection, the harder it is to understand. Meanwhile, as the sheer volume of data grows on the Web, ``storytelling'' is becoming a popular technique in social media for selecting Web resources to support a particular narrative or ``story''.
There are multiple stories that can be generated from an archived collection with different perspectives about the collection. For example, a user may want to see a story that is composed of the key events from a specific Web site, a story that is composed of the key events of the story regardless of the sources, or how a specific event at a specific point in time was covered by different Web sites, etc. In this paper, we provide different case studies for possible types of stories  that can be extracted from a collection. We also provide the definitions and models of these types of stories. 

\end{abstract}


\section{Introduction and Background}
Since it was invented approximately 25 years ago, the Web has developed significantly and new research methodologies have evolved. Moreover, the beginning of Web 2.0 in early 2000 allowed users to contribute born-digital materials to the Web including images, videos, geo-locations, and text. With the emergence of Web 2.0, digital materials have become part of our cultural heritage and preserving the resources of the Web has become essential to facilitate research in history, sociology, political science, media, literature, and other related disciplines.
For many, social media has become the primary resource when an important event occurs \cite{Kwak2010}, and it also can provide the initial spark for important stories (for example, the initial events of the Egyptian Revolution occurred on Facebook \cite{Sutter2, KESSLER, AbigailHauslohner2011}).

\begin{figure*}
\centering 
	\includegraphics[width=0.9\textwidth]{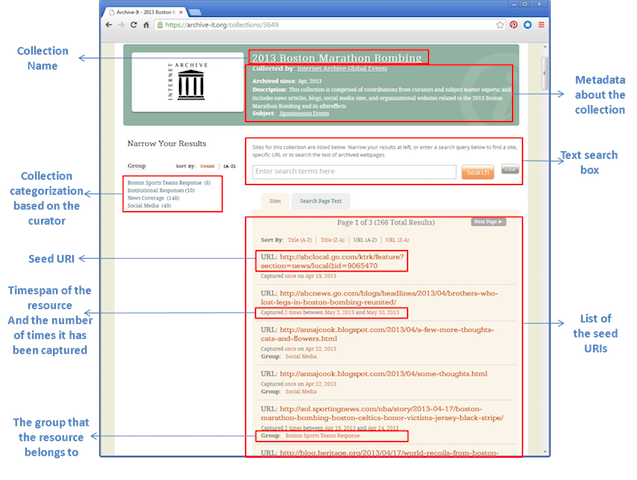}
	\caption{The Archive-It interface of the 2013 Boston Marathon Bombing collection is a list of URIs that are ordered alphabetically.}
	\label{fig:archive-it_anatomy}
\end{figure*}

With the extensive growth of the Web, multiple Web archiving initiatives have been started to archive different aspects of the Web \cite{webarchivesurvey2014}. This was followed by emerging practices and technologies from the archiving communities. For example, the Internet Archive\footnote{\url{http://archive.org/}} (IA), which has been archiving the Web since 1996, generated standards, tools, and technologies to capture Web pages and replay them (e.g., the Wayback Machine \cite{wayback:billion}). Several universities built their own Web archives for research purposes (e.g., the Stanford WebBase Archive) \cite{cho2006stanford}. 

Additionally, multiple archiving initiatives exist to allow people to archive Web resources into themed collections to ensure these resources are available for posterity. 
For example, Archive-It\footnote{\url{http://www.archive-it.org/}}, a subscription service from the IA, allows institutions to develop, curate, and preserve topic-oriented collections of Web resources by specifying a set of seeds, Uniform Resource Identifiers (URIs), that should be crawled periodically. Archive-It provides a listing of all seeds in the collection along with the number of times and dates over which each page was archived, as well as a full-text search of archived pages (Figure \ref{fig:archive-it_anatomy}). Although Archive-It provides users with tools for searching and browsing collections, it is not easy for users to understand the essence of these collections \cite{Padia2012a}.

This paper investigates and describes the abstract model for generating stories from archived collections, along with the terminology and definitions for an archived collection. These stories templates will be used  to automatically extract stories from archived collections \cite{AlNoamany17stories}. 
%
%

\begin{figure*}
\centering 
	\includegraphics[width=1\textwidth]{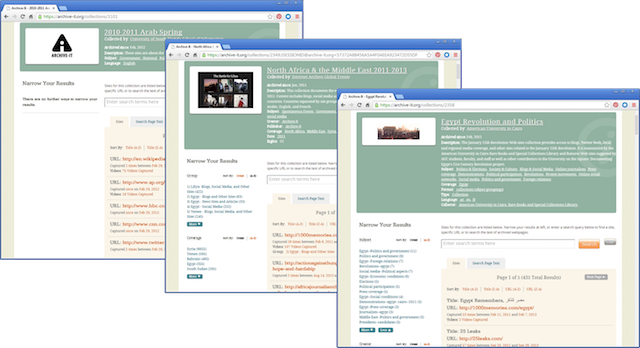}
	\caption{There are multiple collections in Archive-It about the Jan.\@ 25 Egyptian Revolution.}	
	\label{fig:multiple_col_egy}
\end{figure*}


\begin{figure*}
\centering
\subfigure[Archival metadata for the collection.]{\label{fig:egy_browse1}\includegraphics[width=0.47\linewidth]{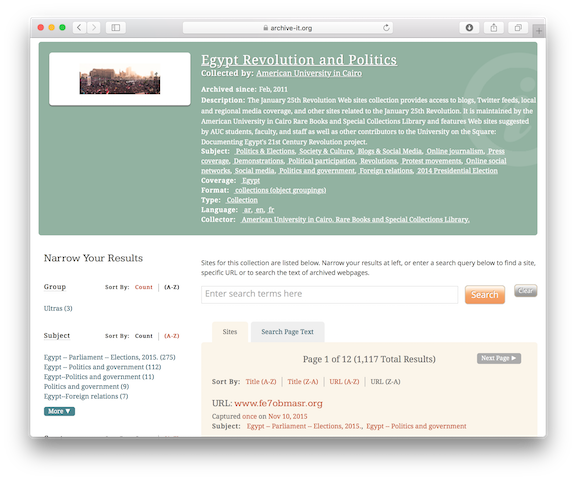}
}
\subfigure[Alphabetical list of URIs in the collection.]{\label{fig:egy_browse2}\includegraphics[width=0.47\linewidth]{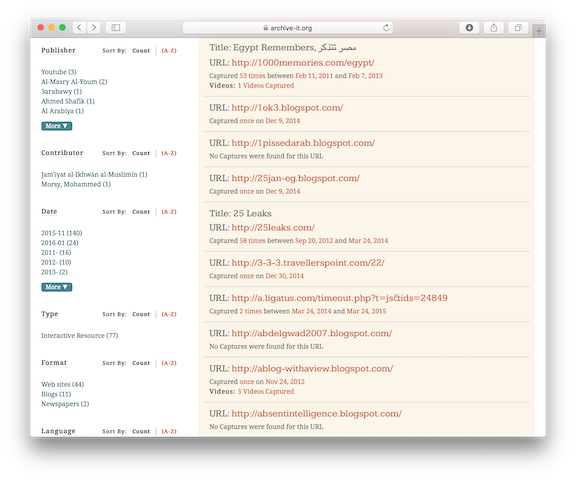}
}
\subfigure[Archived copies of a URI in the collection.]{\label{fig:egy_browse3}\includegraphics[width=0.47\linewidth]{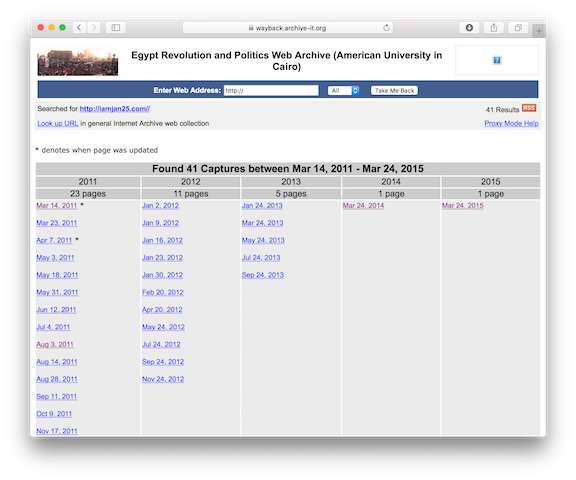}
}
\subfigure[A copy of ``Iam25Jan'']{\label{fig:egy_browse4}\includegraphics[width=0.47\linewidth]{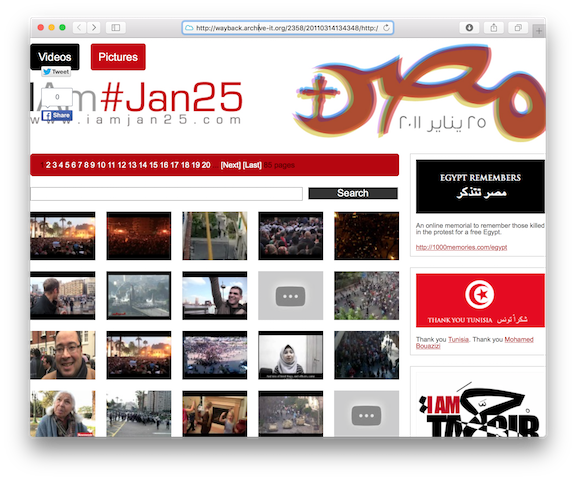}
}
\caption{Current browsing and searching services for the ``Egypt Revolution and Politics'' collection in Archive-It.}
\label{fig:col_browse}
\end{figure*}

\begin{figure*}[!ht]
\centering
\subfigure[Archival metadata for the collection.]{\label{fig:human-rights-1}\includegraphics[width=0.45\linewidth]{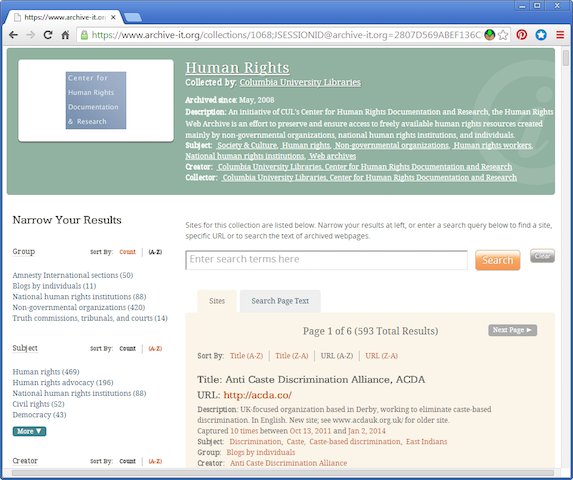}
}
\subfigure[Alphabetical list of URIs in the collection.]{\label{fig:human-rights-2}\includegraphics[width=0.45\linewidth]{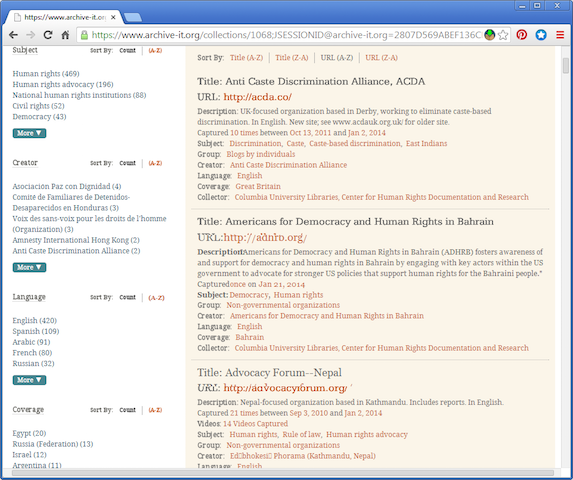}
}
\subfigure[Archived copies of the first URI in the collection.]{\label{fig:human-rights-3}\includegraphics[width=0.47\linewidth]{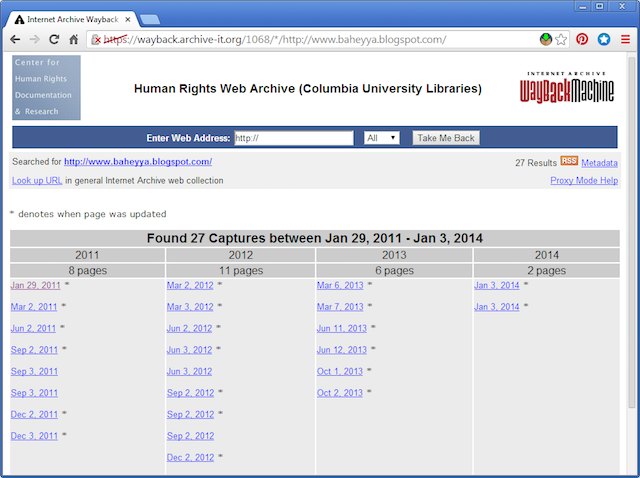}
}
\caption{Current browsing and searching services for the ``Human Rights'' collection in Archive-It.}
\label{fig:col-understanding}
\end{figure*}

\subsection{Collection Understanding}
I want my son, who is now 8 years old, to know what happened during the Jan.\@ 25 Egyptian Revolution as I saw it happening on the Web. Let us assume that he knows about the archived collections that are devoted of archiving important events, such as those at Archive-It. He will use the current browsing interface of \url{archive-it.org} to look for collections related to the Egypt Revolution. 
If he uses the searching and browsing tools that Archive-It provides, he will find about four or five collections containing information about the Jan.\@ 25 Egyptian Revolution (Figure \ref{fig:multiple_col_egy}). 
Aside from the brief metadata about the collection (Figure \ref{fig:egy_browse1}), the interface mainly consists of a list of seed URIs in alphabetical order (Figure \ref{fig:egy_browse2}), and for each of these URIs a list of the times when
the page was archived (Figure \ref{fig:human-rights-3}).
It is not feasible for him to figure out what is inside the collection without going through all the URIs in the collection and their relative archived copies. Understanding the essence of the collection from the current interface of Archive-It is not easy.


%

\textit{Collection understanding} is defined as gaining a comprehensive view of a collection  \cite{Chang2004}. When an archivist creates a collection, it can include 1000s of seed URIs. Over time, each of these URIs can be crawled 100s or 1000s of times, resulting in a collection having thousands to millions of archived Web pages. Understanding the contents and boundaries of a collection is then difficult for most people, resulting in the paradox of the larger the collection, the harder it is to use.

Figure \ref{fig:col-understanding} is another example that shows the current browsing interface for a collection about human rights. It is difficult for users arriving at the page shown in Figure
\ref{fig:human-rights-1} to understand what is in this collection and how it
differs from the approximately $17$ other collections in Archive-It that are also about
human rights, albeit each with their own specialization.

Providing a summary of the content of archived collections is a challenge because there are two dimensions that should be summarized: the URIs that comprise the collection (e.g., \url{cnn.com}) and the archived copies (called ``mementos'') of those URIs at different times (e.g., \url{cnn.com}@$t_1$, \url{cnn.com}@$t_2$,.., \url{bbc.co.uk}@$t_n$). 
Either dimension by itself is difficult, but combined they present a number of challenges, and are hard to adapt to most conventional visualization techniques. 

We have explored applying well-known, advanced visual interfaces (e.g., timelines, wordles, bubble charts, image plots with histogram) to Archive-It collections and the results are sufficient for those already with an understanding of what is in the collection, but they do not facilitate an understanding to those who are unfamiliar with collection \cite{Padia2012a}. 
One problem with the above approaches is there is not an emphasis on ignoring content: there is often an implicit assumption that everything in a collection is equally valuable and should be visualized. Some of the web pages change frequently and some are near-duplicates. Some go off-topic and no longer contribute to the collection. Furthermore, collections grow quickly: the Human Rights collection in Figure \ref{fig:human-rights-1} has nearly 1000 seed URIs, and each URI has between one and 60 archived pages. Visualization techniques with an emphasis on recall (i.e., ``here's everything in the collection'') do not scale. 
\subsection{Memento Terminology}
Memento \cite{VandeSompel2011} is an HTTP protocol extension which enables time travel on the Web by linking the current resources with their prior state. Memento defines the following terms:
\begin{itemize}
 \item URI-R identifies the original resource. It is the resource as it used to appear on the live Web. A URI-R may have 0 or more mementos (URI-Ms).
\item URI-M identifies an archived snapshot of the URI-R at a specific datetime, which is called Memento-Datetime, e.g., URI-M$_{i}$= $URI\mhyphen R @ t_{i}$.
\item URI-T identifies a TimeMap, a resource that provides a list of mementos (URI-Ms) for a URI-R with their Memento-Datetimes 
 \end{itemize}
\section{Living the Past}\label{sec:stories_into}
There are multiple stories that can be generated from an archived collection with different perspectives about the collection. For example, a user may want to see a story that is composed of the key events from a specific Web site, a story that is composed of the key events of the story regardless of the sources, or how a specific event at a specific point in time was covered by different Web sites, etc. We will explore many different types of stories. The story types will be defined and explained in Section \ref{sec:stories_types}.

In the following scenarios, we show manually created stories that bring insight into Archive-It's ``Egypt Revolution and Politics''\footnote{\url{https://archive-it.org/collections/2358/}}, using different sets of archived pages from the collections.

\begin{figure*}
\centering 
	\includegraphics[width=0.7\textwidth]{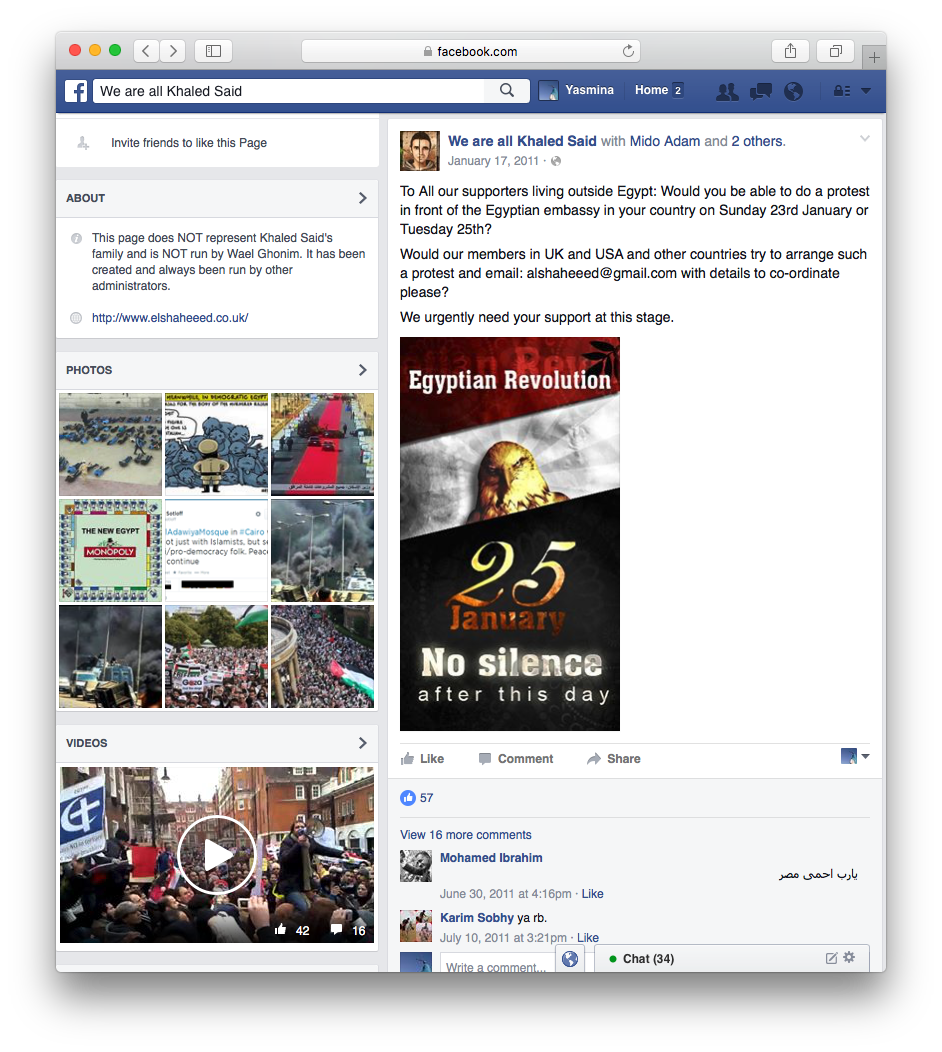}
	\caption{The beginning of the events for the  Jan.\@ 25 Egyptian Revolution started on ``We are All Khaled Saeed'' Facebook page, which formed in the aftermath of Saeed's beating and death. This post is from Jan. 17, 2011 before the start of the Revolution.}	
	\label{fig:egy_rev_start}
\end{figure*}

\subsection{The Jan. 25 Egyptian Revolution}
I was in Norfolk, Virginia when the uprisings of the Jan.\@ 25 Egyptian Revolution started. 
I remember my feeling at that time and how I badly wanted to go back to Egypt and do something for freedom and dignity. I could not do something during the 18 days except watch all the news and social media channels, witnessing the events. It started with a group of brave young Egyptians calling for demonstrations on Facebook and Twitter (Figure \ref{fig:egy_rev_start}).
Millions of people took to the streets in a nationwide protest against President Hosni Mubarak. They aimed to battle injustice, corruption, and poverty. 
Street demonstrations quickly grew into a national revolutionary movement that in 18 days removed Mubarak and his National Democratic Party (NDP). In the following subsections, we will go back in time and see the 18 days and other perspectives about the Egyptian Revolution as they appeared on the Web.

\begin{figure*}
\centering 
	\subfigure[25 Jan.\@ 2011]{
	\includegraphics{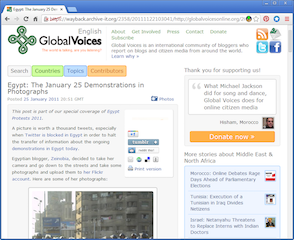}
	\label{fig:egy1}
	}
	\subfigure[27 Jan.\@ 2011]{ 
	\includegraphics{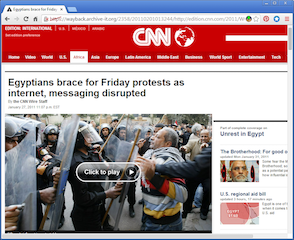}
	\label{fig:egy2}
	}
	\subfigure[31 Jan.\@ 2011]{
	\includegraphics{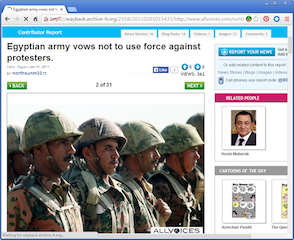}
	\label{fig:egy3}
	}
	\subfigure[31 Jan.\@ 2011]{
	\includegraphics{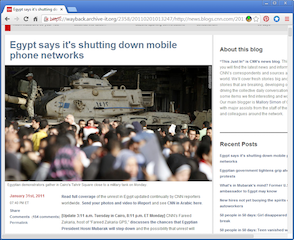}
	\label{fig:egy4}
	}
	\subfigure[01 Feb.\@ 2011]{
	\includegraphics{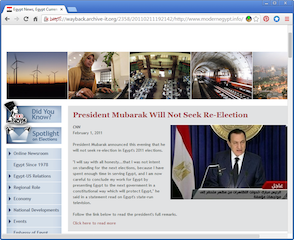}
	\label{fig:egy6}
	}
	\subfigure[02 Feb.\@ 2011]{
	\includegraphics{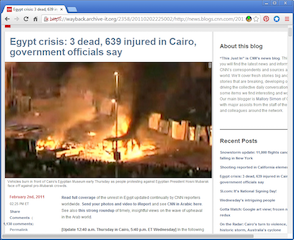}
	\label{fig:egy7}
	}	
	\caption{Coverage of the Egyptian Revolution from different Web sites at different times.}
	\label{fig:s1_egy}
\end{figure*}

\begin{figure*}
\centering 
	\subfigure[02 Feb.\@ 2011]{
	\includegraphics{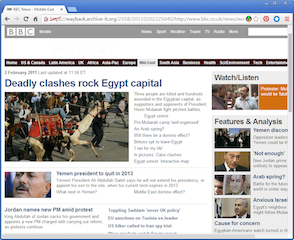}
	\label{fig:egy1-1}
	}
	\subfigure[04 Feb.\@ 2011]{ 
	\includegraphics{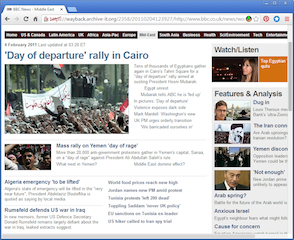}
	\label{fig:egy1-2}
	}
	\subfigure[07 Feb.\@ 2011]{
	\includegraphics{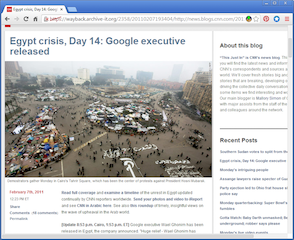}
	\label{fig:egy1-3}
	}
	\subfigure[10 Feb.\@ 2011]{
	\includegraphics{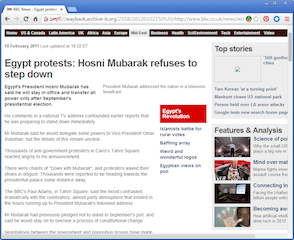}
	\label{fig:egy1-4}
	}
	\subfigure[11 Feb.\@ 2011]{
	\includegraphics{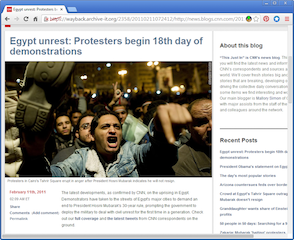}
	\label{fig:egy1-5}
	}
	\subfigure[11 Feb.\@ 2011]{
	\includegraphics{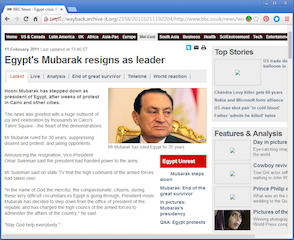}
	\label{fig:egy1-6}
	}	
	\caption{Coverage of the Egyptian Revolution from different Web sites at different times (continued).}
	\label{fig:s1_egy2}
\end{figure*}

\begin{figure*}
\centering 
	\subfigure[2 Feb.\@ 2011]{
	\includegraphics{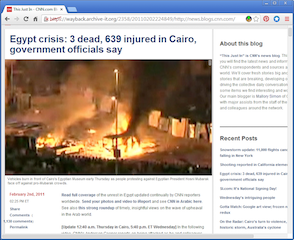}
	\label{fig:egy1cnn}
	}
	\subfigure[4 Feb.\@ 2011]{ 
	\includegraphics{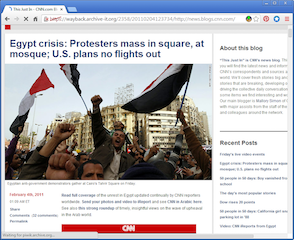}
	\label{fig:egy2cnn}
	}
	\subfigure[5 Feb.\@ 2011]{
	\includegraphics{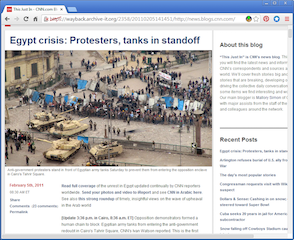}
	\label{fig:egy3cnn}
	}
	\subfigure[7 Feb.\@ 2011]{
	\includegraphics{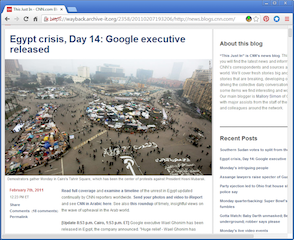}
	\label{fig:egy4cnn}
	}
	\subfigure[11 Feb.\@ 2011]{
	\includegraphics{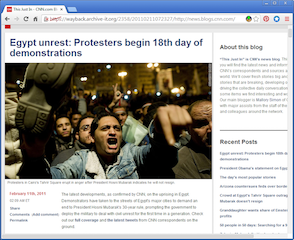}
	\label{fig:egy5cnn}
	}
	\subfigure[11 Feb.\@ 2011]{
	\includegraphics{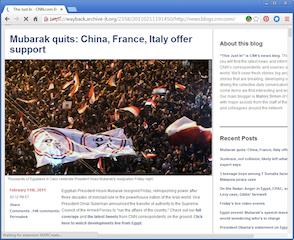}
	\label{fig:egy6cnn}
	}	
	\caption{Coverage of the Egyptian Revolution from CNN's ``This Just In'' blog at different times.}
	\label{fig:s2_sameuri_egy}
\end{figure*}

\subsubsection*{How did the Jan. 25 Egyptian Revolution evolve over time in 18 days?}
\begin{itemize}
\item 2011-01-25: Tens of thousands of young people gathered in Tahrir Square on January 25, 2011 protesting against the government (Figure \ref{fig:egy1}).
\item 2011-01-27: Newspapers started full coverage of the protests with increasing number of protesters because of violent clashes between security forces and protesters (Figures \ref{fig:egy2}).
\item 2011-01-31: The Egyptian military took to the streets, but vowed not to use force against protesters (Figure \ref{fig:egy3}).
\item 2011-01-31 to 2011-02-02: With the increasing anger and the number of protests all over Egypt, the government used multiple ways to stop the protests, such as shutting down access to the Internet \cite{Egy_internet_shutdown1, Egy_internet_shutdown2} and suppressing the media to close communications as these were the main methods that gathered and connected the people. During this period, there was also a speech from Mubarak promising not to seek re-elections, police brutality against the protesters, deadly attacks and clashes from the pro-Mubarak protesters, etc. (Figures \ref{fig:egy4} - \ref{fig:egy7} and Figure \ref{fig:egy1-1}).
\item 2011-02-04: After the number of martyrs increased, the people's anger grew and the numbers of protesters increased significantly all over the country (Figure \ref{fig:egy1-2}). 
\item 2011-02-07: During the protests, Google executive Wael Ghonim revealed that he was behind the account of ``We are All Khaled Saeed''\footnote{\url{https://www.facebook.com/elshaheeed.co.uk/}} Facebook page, which started the anti-government protests that began on Jan.\@ 25. He was arrested during the protests, then he was released (Figure \ref{fig:egy1-3}). 
\item 2011-02-10: Mubarak re-appeared on television in Feb. 10, 2011 and struck a defiant tone (Figure \ref{fig:egy1-4}). 
\item 2011-02-11: The crowd raised their shoes in a response to his speech and insisted that they will not leave until he leaves (Figure \ref{fig:egy1-5}).
\item 2011-02-11: On the Friday of departure (as it was called by the protesters), Egypt's Vice President Omar Suleiman announced that Mubarak would step down after 30 years of rule in an address on state television (Figure \ref{fig:egy1-6}). 
\end{itemize}


By looking at the Web pages in the previous example, the user can get an idea about the Egyptian Revolution's main events from the start of the protests on Jan.\@ 25, 2011 until Mubarak resigned on Feb.\@ 11, 2011. The story in this section is composed of different Web sites at different times.

\subsubsection*{How did CNN cover the 18 days of the 25 Jan. Egyptian Revolution?}
Figure \ref{fig:s2_sameuri_egy} contains different snapshots of the timeline of the Egyptian Revolution as it appeared on \url{http://news.blogs.cnn.com/}. We notice that the start date of the crawl for the URIs in the Egyptian Revolution collection is Feb.\@ 1, 2011, which is seven days after the start date of the Egyptian Revolution (Jan.\@ 25, 2011). 

This scenario shows the evolution of a single page through time. There are several cases where the user might want to see the evolution of a single Web page through time \cite{Jatowt2004}. For example, a user might be interested in the main changes of a popular Web site, or key events from specific Web sites during given period. 

\begin{figure*}[t]
\centering 
	\includegraphics[width=0.6\textwidth]{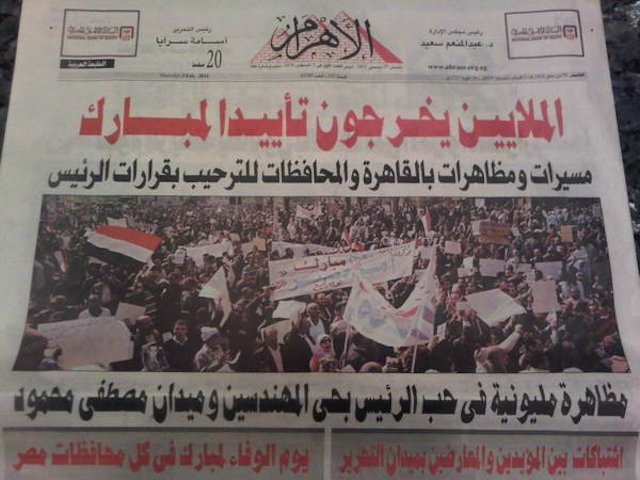}
	\caption{Egyptian State Newspaper, \textit{Al Ahram}: ``Millions go out in support of Mubarak: Demonstrations in Cairo and surrounding areas to welcome - Mubarak's latest decisions,  Millions demonstrate for their love of the president in Muhandiseen and Mustafa Mahmood Square''. Source: \protect\url{http://imgur.com/DbtK1}}
	\label{fig:fake_new}
\end{figure*}

\subsubsection{How did different newspapers cover Mubarak resigning?}
Egypt is the largest Arabic country and has played a central role in Middle Eastern politics. Therefore, there were widely varying reactions toward the Egyptian Revolution, nationally and internationally. 
This is how Pasha described the coverage of the Egyptian media during the 18 days of demonstrations \cite{pasha2011islamists}:
\begin{quote}
Egyptian media, including \textit{Al-Ahram}, falls under the authoritarian type, where the ruling
regime and the elites monopolize media outlets. The authoritarian type indicates that journalism is
subservient to the interests of the state in maintaining social order and achieving political goals. Saying
that \textit{Al-Ahram} is under the authoritarian type implies it avoids criticism to the President, the government
policies or officials, and it censors publishing any material that challenges the established order.
\end{quote}

Inside Egypt, the official newspapers did not cover the protests as they were happening. They were biased against the protests and supported Mubarak until he stepped down \cite{egy_media_lie_artcile}. An example shown in Figure \ref{fig:fake_new} contains the headline from Feb.\@ 3, 2011 on the cover page of \textit{Al-Ahram}, the most widely circulating state-owned daily newspaper and the second oldest newspaper in Egypt, founded in 1875. It reads ``Millions go out in support of Mubarak'' and has no news about the protests against Mubarak at that time.

A wide range of research has been conducted to study the media's coverage of the Egyptian Revolution \cite{ghobrial2014politics, youssef2012critical, pasha2011islamists}. These studies discovered that the coverage by the governmental newspapers of the Egyptian demonstration differed from the international newspapers. Youssef Ahmed presented many examples for how \textit{Al-Ahram} was prone to accentuate protesters' acts of violence and published many articles to affect people's opinions against the protests  \cite{youssef2012critical}.

\begin{figure*}
\centering 
	\subfigure[11 Feb.\@ 2011]{
	\includegraphics[width=0.4\textwidth]{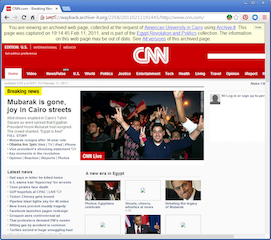}
	\label{fig:egy3-1}
	}
	\subfigure[11 Feb.\@ 2011]{ 
	\includegraphics[width=0.4\textwidth]{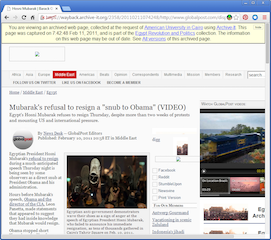}
	\label{fig:egy3-2}
	}
	\subfigure[11 Feb.\@ 2011]{
	\includegraphics[width=0.4\textwidth]{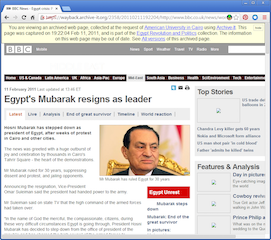}
	\label{fig:egy3-3}
	}
	\subfigure[11 Feb.\@ 2011]{
	\includegraphics[width=0.4\textwidth]{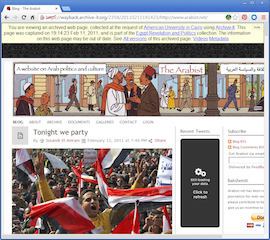}
	\label{fig:egy3-4}
	}
	\subfigure[11 Feb.\@ 2011]{
	\includegraphics[width=0.4\textwidth]{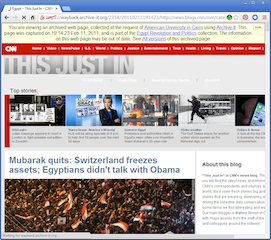}
	\label{fig:egy3-6}
	}
	\subfigure[11 Feb.\@ 2011]{
	\includegraphics[width=0.4\textwidth]{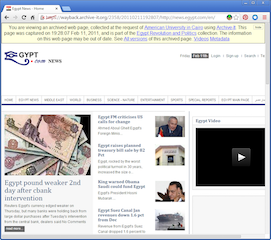}
	\label{fig:egy3-7}
	}	
	\caption{Coverage of the Egyptian Revolution from different sites at a specific time (Feb.\@ 11, 2011).}
	\label{fig:s3_samedate_egy}
\end{figure*}

The pages shown in Figure \ref{fig:s3_samedate_egy} cover through multiple sites the reactions to Mubarak stepping down. Figure \ref{fig:egy3-7} shows the coverage of one of the national Egyptian newspaper on Feb.\@ 11, 2011, the day when Mubarak stepped down. Although the page shows the reaction of Saudi Arabia on the Revolution and their support of Mubarak, it does not have any coverage for Mubarak stepping down. 

To gain insight about a specific event, there is a need to know the date of the event. If a user wants to browse all the pages that were crawled on Feb.\@ 11, 2011 (e.g., the pages in Figure \ref{fig:s3_samedate_egy}), there is no way to do this with the current Archive-It interface. 

\section{Conventions and Definitions}\label{sec:definition}
In this section, we give the notions and the conventions we will use for defining the possible generated stories from archived collections. 
It is possible for a collection to be summarized with more than one kind of story (depending on the nature of the collection as well as curator or user preferences). Before specifying the possible types of stories (Section \ref{sec:stories_types}), we first define the archived collections. 

An Archive-It collection $(C)$ is a set of seed URIs collected by the users from the Web $(W)$, where $C \subset W$. Each seed URI has mementos.
  
A collection $C$ can be formally defined as following:
\begin{eqnarray}\label{eq:col_def}
C &=& \{URI \mhyphen T_1,URI \mhyphen T_2,...,URI \mhyphen T_n\}\mbox{ where }  \nonumber \\
    & & URI \mhyphen T = \{URI \mhyphen M_1, URI \mhyphen M_2,...,URI \mhyphen M_x\} \nonumber \\
    & & \mbox{ and } URI \mhyphen M_i \mbox{ is } URI \mhyphen R \MVAt t_i 
\end{eqnarray}

\begin{figure*}
\centering 
	\includegraphics[width=0.8\textwidth]{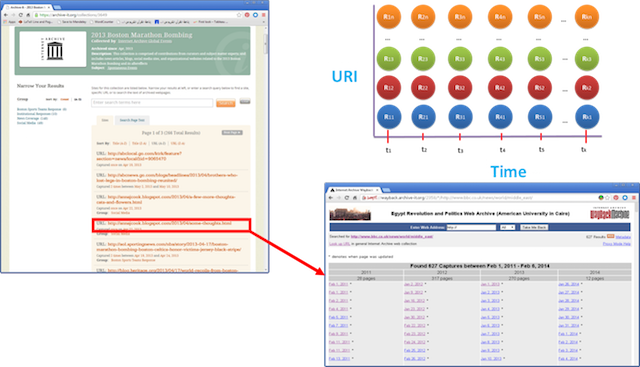}
	\caption{The archived collection has two dimensions: URI and time} 
	\label{fig:col_dim}
\end{figure*}

\begin{figure*}
\centering 
	\subfigure[Fixed-Fixed: Same URI, Same time]{
	\includegraphics[width=0.5\textwidth]{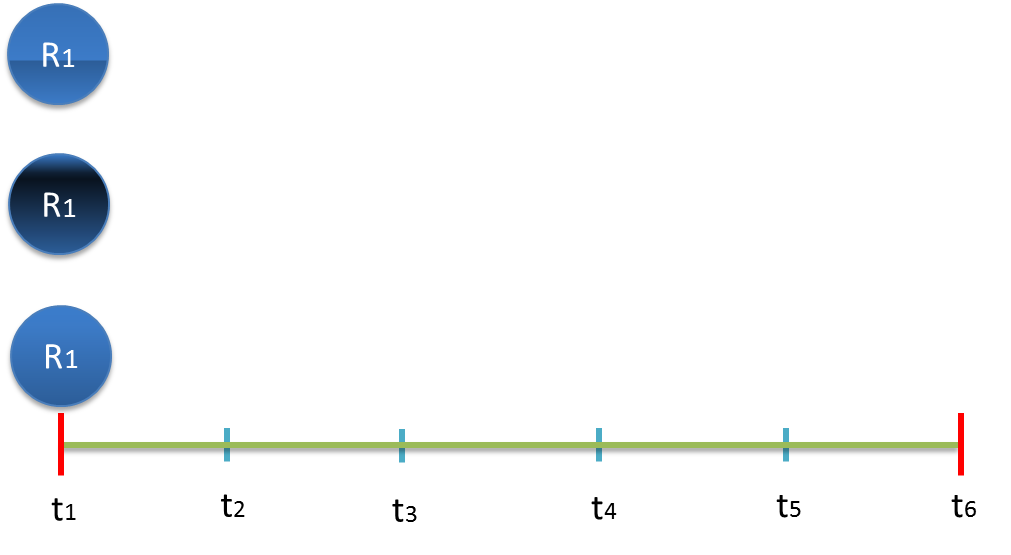}
	\label{fig:story_model0}
	}
	\subfigure[Sliding-Sliding: Different URIs, different times]{
	\includegraphics[width=0.5\textwidth]{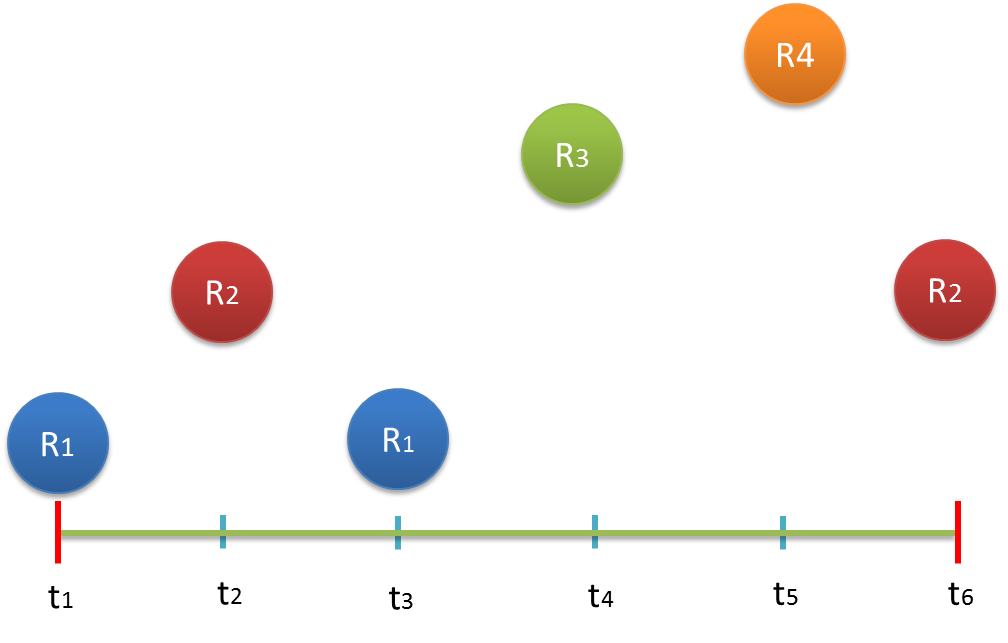}
	\label{fig:story_model1}
	}
	\subfigure[Fixed-Sliding: Same URI-R, different times]{
	\includegraphics[width=0.5\textwidth]{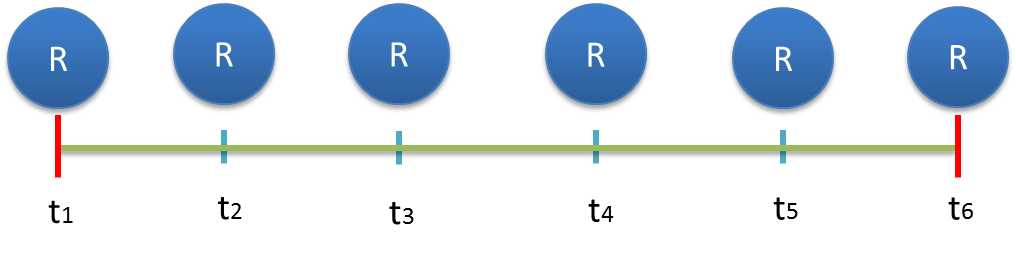}
	\label{fig:story_model2}
	}
	\subfigure[Sliding-Fixed: Different URIs, same time.]{
	\includegraphics[width=0.5\textwidth]{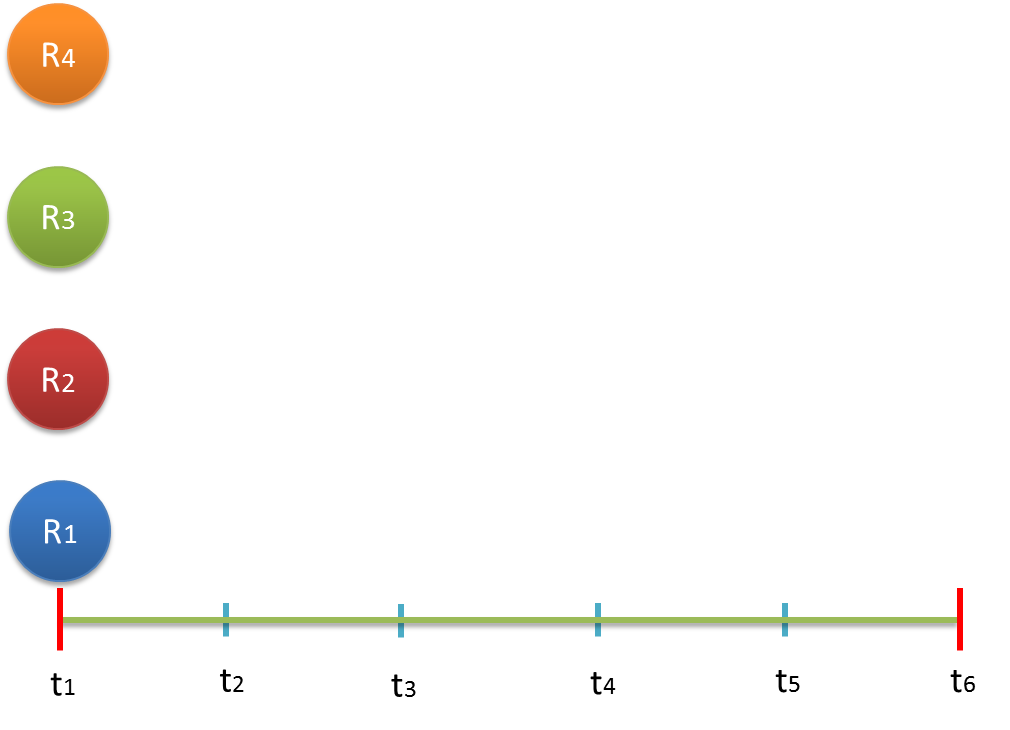}
	\label{fig:story_model3}
	}
		\caption{There are different models for the story that can be created from the collection. The color maps to the unique URI-R.}
	\label{fig:story_model}
\end{figure*}

\section{Types of Stories Generated from Archived Collections}\label{sec:stories_types}
An archived collection has two dimensions. As we mentioned before, the collection is composed of a set of seed URIs and each seed has many copies through time (Figure \ref{fig:col_dim}). There may be multiple stories that convey different perspectives of the collection, such as the examples of Figures \ref{fig:s1_egy} - \ref{fig:s3_samedate_egy}.  We list four possible kinds of stories in Table  \ref{tab:storytypes}. We name each story according to the change that happens to the URI and time. 
It is also possible that there are additional types of stories beyond those in Table \ref{tab:storytypes}, and we plan to investigate this in future work.

\begin{table*}[h]
\centering
\caption{\label{tab:storytypes}Four basic story types (others may be possible).}
\begin{tabular}{crc|c}
& & \multicolumn{2}{c}{Time:} \\
& & fixed & sliding \\
\cline{3-4}
\cline{3-4}
\multirow{2}{*}{URIs:}
& fixed & differences in GeoIP, & evolution of a single page \\
& & mobile, etc. & (or domain) through time \\
\cline{3-4}
& sliding & different perspectives & broadest possible coverage \\
& & at a point in time & of a collection \\
\cline{3-4}
\end{tabular}
\end{table*}

We present the definition for each story below, along with a model in Figure \ref{fig:story_model}. The different colors in Figure \ref{fig:story_model} map to different URI-Rs. We use Memento terminology (URI-T, URI-M, and URI-R) in the definitions.

\begin{figure*}[t]
\centering
	\subfigure[The cnn.com memento when crawled with a desktop Mozilla user-agent accessed from a Mac.]{
	\includegraphics[width=0.46\textwidth]{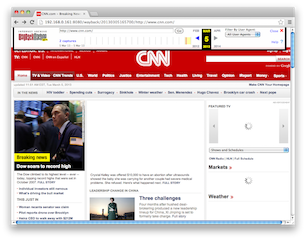}
	\label{fig:s0_1}
	}
	\subfigure[The cnn.com memento when crawled with an iPhone mozilla user-agent accessed from a Mac.]{
	\includegraphics[width=0.46\textwidth]{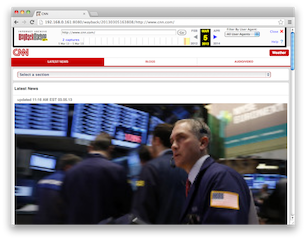}
	\label{fig:s0_2}
	}
	\caption{Mementos differ based on the parameters influencing the representations at crawl/capture time and the devices used to access the mementos \cite{kellybrunelledlib}.}
	\label{fig:diff_rep}
\end{figure*}

\subsection{Fixed Page, Fixed Time}
Fixed Page, Fixed Time (FPFT) is defined as a different representation for the same Web site because of GeoIP, mobile, and other environmental factors (e.g., Figure \ref{fig:diff_rep}) \cite{kellybrunelledlib}. It is generated using the same URI at a specific point of time with differences in the representation. The model for this story is shown in Figure \ref{fig:story_model0}.

\begin{eqnarray} 
\mbox{FPFT} &=& (URI \mhyphen M_{i}, URI \mhyphen M_{i}, ..., URI \mhyphen M_{i}),\mbox{  where }  \nonumber \\
  & & URI \mhyphen M_{i}= URI \mhyphen R \MVAt t_i
\end{eqnarray}

\subsection{Sliding Page, Sliding Time} 
Sliding Page, Sliding Time (SPST) is defined as the broadest possible coverage of a collection. It is generated using different URIs at different times.

\begin{small}
\begin{eqnarray} 
\mbox{SPST} &=& (URI \mhyphen M_{1}, URI \mhyphen M_{2}, ..., URI \mhyphen M_{k}),\mbox{  where }  \nonumber \\
  & & URI \mhyphen M_{i}= URI \mhyphen R \MVAt t_i\mbox{ and }t_i \neq t_j
\end{eqnarray}
\end{small}

\subsection{Fixed Page, Sliding Time} 
Fixed Page, Sliding Time (FPST) is defined as the evolution of a single page (or domain) through time Figure (\ref{fig:story_model3}). The possible scenario of this story is when a user wants to see how the story evolved over time from a specific Web site, e.g., \url{cnn.com}. 
\begin{small}
\begin{eqnarray} 
\mbox{FPST} &=& (URI \mhyphen M_{1}, URI \mhyphen M_{2}, ..., URI \mhyphen M_{k}),\mbox{  where }   \nonumber \\
  & & URI \mhyphen R(URI \mhyphen M_i) = URI \mhyphen R(URI \mhyphen M_j)\mbox{ and } \nonumber \\
  & & URI \mhyphen M_{i}= URI \mhyphen R \MVAt t_i 
\end{eqnarray}
\end{small}

\subsection{Sliding Page, Fixed Time} 
Sliding Page, Fixed Time (SPFT) is defined as different perspectives at a point in time. It is generated using different URI-Rs at nearly the same datetime.

\begin{small}
\begin{eqnarray} 
\mbox{SPFT} &=& (URI \mhyphen M_{1}, URI \mhyphen M_{2}, ..., URI \mhyphen M_{k}),\mbox{  where }   \nonumber \\
  & & URI \mhyphen R(URI \mhyphen M_i) \neq URI \mhyphen R(URI \mhyphen M_j)\mbox{  and }
  \nonumber \\ 
  & & URI \mhyphen M_{i}= URI \mhyphen R \MVAt t_i
\end{eqnarray}
\end{small}
 
Note that the Fixed-Fixed story can not be supported by the current capabilities of Web archives \cite{kellybrunelledlib}. While Heritrix provides archivists the ability to modify the user-agent string to crawl different representations, such as mobile Web, archives currently do not provide users the ability to navigate representations by their environmental influences. 
Kelly et al. \cite{kellybrunelledlib} proposed a method for identifying personalized representations in Web archives through a modified Wayback Machine to add environmental dimensions to browsing the past Web.


\section{Conclusions and Future Work}
Archiving Web pages into themed collections is a method for ensuring these resources are available for posterity. Many institutions archive the Web, resulting in tremendous amount of archive pages that have thousands of mementos.  There are multiple stories that can be generated from an archived collection with different perspectives about the collection depending on the nature of the collection as well as the curators' preference. 

In this paper, we provided conceptual models and definitions for possible types of stories that can be generated from archived collections: Fixed Page, Fixed Time (FPFT), a single page at a specific time; Sliding Pages, Sliding Time (SPST), a broad summary of different URIs through time that provides an overview of the collection from different Web sites; Sliding Page, Fixed Time (SPFT), different URIs at nearly the same time that provide different perspectives at a point in time; Fixed Page, Sliding Time (FPST), same Web site at different times that provides an evolution of a single page (or domain) through time. 
Note that with the current capabilities of Web archives, the Fixed Page, Fixed Time (FPFT) story cannot be supported because archives currently do not provide users with the ability  to navigate representations by their environmental influences \cite{kellybrunelledlib}. We also provided different scenarios that shows the need of these types for humanities, journalists, and social science researchers.

We provided preliminary investigation for possible types of stories generated from archived collections. There might be other types of stories that can be generated. We plan to collaborate with humanities researchers to conduct user studies on important events, e.g., the Arab Spring, and check if there are other possible types of stories and if a specific kind of story provides the best insight into the
events and the corresponding collections. For example, how do the Sliding Page, Fixed Time stories help humanities researchers to get different perspectives about news coverage and how much time is saved from manual search by providing them this kind of story.

\section{Acknowledgments}
This work supported in part by the Institute Museum and Library Services (LG-71-15-0077-15).  We thank the Archive-It team and partners for creating the gold standard data set.

%
%
\bibliographystyle{abbrv}
\bibliography{alnoamany17} 
\balancecolumns
\end{document}